\documentclass{proc} 
\usepackage{edbkps} 
\let\footnote\savefootnote
\let\footnotetext\savefootnotetext
\let\footnoterule\savefootnoterule
\normallatexbib 

\usepackage{t1enc}
\usepackage{amssymb}
\usepackage{amsmath}

\usepackage[dvips]{graphicx}

\begin{document}

\articletitle[A Clifford Algebra approach to the Discretizable Molecular Distance Geometry Problem]
{A Clifford Algebra approach to the\\
Discretizable Molecular Distance Geometry Problem
}

\author{Alessandro Andrioni\altaffilmark{1}}

\altaffiltext{1}{IMECC, University of Campinas,
Campinas,
Brazil,
\email{andrioni@member.ams.org}}


\anxx{Alessandro Andrioni, University of Campinas\,
Campinas\,
Brazil\, \emailfont{andrioni@member.ams.org}}

\begin{abstract}
The Discretizable Molecular Distance Geometry Problem (DMDGP) consists in a subclass of the Molecular Distance Geometry Problem for which an embedding in ${\mathbb{R}^3}$ can be found using a Branch \& Prune (BP) algorithm in a discrete search space. We propose a Clifford Algebra model of the DMDGP with an accompanying version of the BP algorithm.
\end{abstract}


\section{The discretizable molecular distance geometry problem}
\label{dmdgp}
The molecular distance geometry problem (MDGP) consists in finding coordinates in a three-dimensional space of a set of points $\{x_1, x_2, \dots, x_n \}$ for which some of the Euclidean distances between them are known~\cite{crippen}. Let $G = (V, E, d)$ be a simple weighted undirected graph where each vertex in $V$ corresponds to a point in $\mathbb{R}^3$, and the weight of an edge corresponds to the distance $d$ between the respective points. Formally, the MDGP can be defined as follows~\cite{dmdgp}:
\begin{definition}
(MDGP). Let $G = (V, E, d)$ be a simple weighted undirected graph. The MDGP is the problem of finding a function
\[ x : V \to \mathbb{R}^3 \]
such that
\[ \forall (u,v) \in E, \ ||x_u-x_v|| = d_{uv}, \]
where $x_u = x(u)$ and $x_v = x(v)$.
\end{definition}

The discretizable molecular distance geometry problem (DMDGP) is a subset of the MDGP, but with two extra assumptions~\cite{dmdgp}:
\begin{definition}
(DMDGP). Let $G = (V, E, d)$ be a simple weighted undirected graph associated to an instance of the MDGP. Let us suppose that there is a \textit{total order relation} on the vertices of $V$. The DMDGP consists in all the instances of the MDGP satisfying the following two assumptions:
\begin{enumerate}
\item $E$ contains all cliques on quadruplets of consecutive vertices;
\item the following strict triangular inequality must hold:
\[ \forall v \in \{ 1, \dots, n-2 \},  \ d_{v,v+2} < d_{v,v+1} + d_{v+1,v+2}, \]
where $n$ is the number of vertices in $V$.
\end{enumerate}
\end{definition}

The reasons why these assumptions are useful and realistic is outside of the scope of this work, and is extensively addressed in \cite{dmdgp} and \cite{survey}, but they allow us to discretize the problem in the following manner: 
suppose we have three points in $\mathbb{R}^3$, and the distances to a fourth point.
We can then construct three spheres, each one centered in one of the points and with a radius of its distance to the fourth one. These three spheres have an intersection characterized by two points (with probability 1, thanks to the assumptions above~\cite{survey}): the two possibilities for the fourth point.
However, if we have additional information, we can decide whether one of them is invalid or not.
Repeating this process, we have at most $2^{n-3}$ ways to position $n$ points (up to rotation and translation).

The Branch \& Prune (BP) is an algorithm proposed in \cite{itor2008} to solve the DMDGP by exploiting this discretization.
The BP was further developed~\cite{symmbp} to use another characteristic of the discretization, namely that when constructing a new point, the two possibilities are symmetric in regard to the plane determined by the three previous points. This means we can construct other realizations from an initial one by knowing just the different branches taken and then applying reflections through the right planes.

\section{Clifford algebra}
Clifford algebras are a refinement of both the Hamilton quaternions and the extensive algebra of Hermann Grassmann~\cite{grassmann} condensed in one structure.
Clifford himself called his work \textit{geometric algebra}~\cite{clifford}, but the term most commonly used now is Clifford algebra. His work had a geometric flavor, and was heavily explored by both mathematicians and physicists, including \'Elie Cartan and Paul Dirac, usually in the context of differential geometry and quantum mechanics~\cite{Dorst2007}.

A revival of the use of real Clifford algebras for geometric purposes was spearheaded by David Hestenes~\cite{Hestenes87}, and culminated in the modern geometric algebra and its operational models of Euclidean geometry, including conformal geometric algebra (CGA)~\cite{Hestenes01oldwine}.
CGA allows a rich representation of Euclidean motions in a coordinate-free manner, and the link between distance geometry and conformal geometric algebra was already studied by Dress and Havel~\cite{dresshavel}.

We try to follow the notation and the formulation introduced in \cite{Dorst2007}, and recommend it as a good introduction to the subject, but we give a summary of central Clifford algebra ideas used in this work.

There are two main products which are used: the geometric (or Clifford) product, and the outer (or wedge) product\footnote{The geometric product of $a$ and $b$ is denoted by $ab$, and their outer product by $a \wedge b$.}. Both of them algebraically encode the idea of working with \textit{oriented subspaces} of a vector space, allowing a \textit{``multivector''} representation of points, lines, planes and hyperplanes. An interesting fact is that the geometric product allows the \textit{inverse} of some multivectors to be defined, and thus it permits the representation of orthogonal or even conformal (in CGA) transformations using an object called versor.

A versor is the result of a multiplication of vectors using the geometric product, it is always invertible and it is applied to another multivector by ``sandwiching'', that is, if $V$ is a versor and $A$ a multivector, we can apply the versor by calculating $VAV^{-1}$.
Versors also correctly preserve its underlying geometric structure without need for adaptations, being then a suitable basis for the representation of geometric computations.

%

\begin{definition}
(Conformal geometric algebra of the three-dimensional Euclidean space). The conformal geometric algebra (CGA) of the three-dimensional Euclidean space is an extension of $\mathbb{R}^3$ by means of two extra vectors $e$ and $\bar{e}$, which square respectively to $1$ and $-1$. 
However, it is more convenient to use $\infty = \bar{e} - e$ and $o = \frac{1}{2}(\bar{e} + e)$, both of which square to $0$, and represent a \textit{point at infinity} and a \textit{point at the origin}, respectively. This permits a fully coordinate-free representation of Euclidean geometry, a fact which will be exploited in our algorithm.
\end{definition}

In CGA, versors encode all conformal transformations, including isometries and homotheties. In fact, it is the smallest known model of Euclidean geometry which allows the full representation of Euclidean transformations as versors. For convenience, we introduce special names to two kinds of versors: those which represent rotations (rotors), and those which represent translations (translators).

Composition of rotors and translators is more efficient than that of rotation matrices in up to 10 dimensions and uses less storage in up to 6 dimensions, a good evidence of the appropriateness of using CGA for geometric computing. It is also simple to convert versors to matrices, if the need arises~\cite{Dorst2007}.




%

\section{BP with Clifford algebra}
We assume that our instance is a molecule of $n$ atoms for which we denote the \textit{bond lengths} $d_{i-1, i}$ for $i = 2, \dots, n$, the \textit{bond angles} $\theta_{i-2, i}$ for $i = 3, \dots, n$ and the \textit{dihedral angles} $\omega_{i-3, i}$ for $i = 4, \dots, n$.

The main idea of the BP with Clifford Algebra is to use \textit{rotors} and \textit{translators} to represent the calculation of a point from its predecessors, instead of transformation matrices based on homogeneous coordinates~\cite{dmdgp}. Since a combination of rotor and translator needs at most 8 coordinates to be represented, this represents a memory gain against the traditional $4 \times 4$ matrices.

Another advantage of using CGA to work on the DMDGP is to exploit the inherent symmetries in the problem, since reflection through a plane is a simple operation in CGA represented by a reflection versor, cheapening considerably the cost of calculating alternative conformations.

An algorithm to calculate the two possible points from its predecessors is presented here as ``\textit{\textbf{Algorithm I: Compute the $i$-th points}}''. It creates two rotors: one for the bond angle and one for the dihedral angle, and apply them to a translator to generate $F$, an Euclidean transformation which takes $x_{i-1}$ to $x_i$. Notice in the algorithm the particular way the rotors are constructed, which is reminiscent of Euler's formula and the polar forms of complex numbers or quaternions.
The pruning phase is implemented as in the original BP~\cite{survey}.

A computer implementation of this new version of the BP already exists for the GAViewer software \cite{Dorst2007}, but as the software was not made with efficiency needs in mind, it is only useful as a visualization tool.
A ``production-ready'' implementation using the software Gaigen~\cite{gaigen} capable of using data extracted from the Protein Data Bank is currently being written, and should be complete in time for the DGA2013, along with a performance analysis and a comparison with the existing implementations of the original BP~\cite{itor2008}.
%

\begin{algorithm}
\
{\bit Algorithm I: Compute $i$-th points} ($x_{i-3}, x_{i-2}, x_{i-1}, \theta, \omega, d $):
\ {\bf $\Pi$} := $x_{i-3} \wedge x_{i-2} \wedge x_{i-1} \wedge \infty$;
\ {\bf $R_1$} := $e^{\frac{\theta}{2}((\Pi x_{i-2} \wedge x_{i-1}) \wedge \infty )^* } $; \note{/* Create the bond angle rotor */}
\ {\bf $v$} := $x_{i-1} - x_{i-2}$;
\ {\bf $\omega'$} := $\omega - \frac{\pi}{2}$;
\ {\bf $R_2$} := $e^{\frac{\omega'}{2} (x_{i-2} \wedge x_{i-1} \wedge \infty )^* } $; \note{/* Create the dihedral angle rotor */}
\ {\bf $T$} := $1 - \frac{d}{2} \frac{v}{||v||} \infty$; \note{/* Create the translator */}
\ {\bf $F$} := $R_2 \, R_1 \, T R_1^{-1} R_2^{-1}$; \note{/* Combine two rotors and a translator in one versor */}
\ {\bf $x_i$} := $F \, x_{i-1} F^{-1}$ \note{/* Apply it $x_{i-1}$ */}
\ {\bf $x_i'$} := $\Pi x_i \Pi^{-1}$ \note{/* Reflect $x_i$ */}
\ {\it return} $x_i, x_i'$.
\end{algorithm}

\section{Conclusions and future work}
This is one of the first practical applications of conformal geometric algebra in distance geometry and it shows its excellence in representing complex geometric operations in a simple manner. The subsumption of both quaternions and homogeneous coordinates by CGA allows it to clarify the notation and to better expound inherent geometric properties in problems.

We expect to see more developments in that regard in the future, as more accessible, efficient and high-level implementations of geometric algebra appear, as it is suited for scientific computing and is already being used in the fields of robotics, computer graphics, computer vision and artificial neural networks~\cite{Corrochano2010}~\cite{Dorst2007}~\cite{Perwass}.

An extension of this work to handle generalizations of the DMDGP is expected, as conformal geometric algebra has a great richness in ways of creating and manipulating its objects.
Another possible line of work would be to try to apply CGA techniques to other problems involving molecular symmetries and rotations, such as \cite{Fritzer2001} and \cite{Karney2007}.


\begin{acknowledgments}
The authors would like to thank the Brazilian research agencies FAPESP, CNPq and CAPES for the financial support.
\end{acknowledgments}

\bibliographystyle{plain}

\begin{thebibliography}{10}

\bibitem{Corrochano2010}
E.~Bayro-Corrochano.
\newblock {\em Geometric Computing - for Wavelet Transforms, Robot Vision,
  Learning, Control and Action.}
\newblock Springer, 2010.

\bibitem{clifford}
W.~Clifford.
\newblock Applications of {G}rassmann's extensive algebra.
\newblock {\em American Journal of Mathematics}, 1:350--358, 1878.

\bibitem{grassmann}
J.~Collins.
\newblock An elementary exposition of {G}rassmann's ``{A}usdehnungslehre,'' or
  theory of extension.
\newblock {\em The American Mathematical Monthly}, 6:193--198, 1899.

\bibitem{crippen}
G.~Crippen and T.~Havel.
\newblock {\em Distance Geometry and Molecular Conformation}.
\newblock Wiley, New York, 1988.

\bibitem{Dorst2007}
L.~Dorst, D.~Fontijne, and S.~Mann.
\newblock {\em Geometric Algebra for Computer Science: An Object-Oriented
  Approach to Geometry}.
\newblock Morgan Kaufmann Publishers Inc., San Francisco, 2007.

\bibitem{dresshavel}
A.~Dress and T.~Havel.
\newblock Distance geometry and geometric algebra.
\newblock {\em Foundations of Physics}, 23:1357--1374, 1993.

\bibitem{gaigen}
D.~Fontijne.
\newblock Gaigen 2: a geometric algebra implementation generator.
\newblock In {\em Proceedings of the 5th international conference on Generative
  programming and component engineering}, GPCE '06, New York, 2006. ACM.

\bibitem{Fritzer2001}
H.~Fritzer.
\newblock Molecular symmetry with quaternions.
\newblock {\em Spectrochimica Acta Part A: Molecular and Biomolecular
  Spectroscopy}, 57:1919--1930, 2001.

\bibitem{Hestenes01oldwine}
D.~Hestenes.
\newblock Old wine in new bottles: A new algebraic framework for computational
  geometry.
\newblock In {\em Advances in Geometric Algebra with Applications in Science
  and Engineering}, pages 1--14, 2001.

\bibitem{Hestenes87}
D.~Hestenes and G.~Sobczyk.
\newblock {\em Clifford Algebra to Geometric Calculus: A Unified Language for
  Mathematics and Physics}.
\newblock Fundamental Theories of Physics. Springer, 1987.

\bibitem{Karney2007}
C.~Karney.
\newblock Quaternions in molecular modeling.
\newblock {\em Journal of Molecular Graphics and Modelling}, 25:595--604, 2007.

\bibitem{dmdgp}
C.~Lavor, L.~Liberti, N.~Maculan, and A.~Mucherino.
\newblock The discretizable molecular distance geometry problem.
\newblock {\em Computational Optimization and Applications}, 52:115--146, 2012.

\bibitem{itor2008}
L.~Liberti, C.~Lavor, and N.~Maculan.
\newblock A branch-and-prune algorithm for the molecular distance geometry
  problem.
\newblock {\em International Transactions in Operational Research}, 15:1--17,
  2008.

\bibitem{survey}
L.~Liberti, C.~Lavor, N.~Maculan, and A.~Mucherino.
\newblock Euclidean distance geometry and applications. {Tech. Report}
  q-bio.qm/1205.0349, ar{X}iv, 2012.

\bibitem{symmbp}
A.~Mucherino, C.~Lavor, and L.~Liberti.
\newblock {{E}xploiting symmetry properties of the discretizable molecular
  distance geometry problem}.
\newblock {\em Journal of Bioinformatics and Computational Biology}, 10, 2012.

\bibitem{Perwass}
C.~{Perwass}.
\newblock {\em {Geometric Algebra with Applications in Engineering}}.
\newblock Springer Berlin Heidelberg, 2009.

\end{thebibliography}
\chapbblname{AndrioniDGA}  
\chapbibliography{andrioni}

\end{document}